\documentclass[aps,prb,preprint,a4paper,amsfonts,amssymb,floatfix,amsmath,showpacs,showkeys,unsortedaddress]{revtex4}
\usepackage{natbib}
\usepackage{graphicx}
\usepackage{SIunits}

\newcommand{\mang}[4]{\mathrm{#1_{#2}}\mathrm{#3_{#4}}\mathrm{MnO}_{3}{}}

\newcommand{\tc}{\mathrm{T_{C}}}

\newcommand{\tfmi}{\mathrm{T_{FMI}}}
\newcommand{\rhonot}{\rho_0}

\newcommand{\te}{\mathrm{T_{e}}}
\newcommand{\tph}{\mathrm{T_{ph}}}

\newcommand{\ef}{\mathrm{E_{F}}}
\newcommand{\tnot}{\mathrm{T_{0}}}

\newcommand{\deltacg}{\Delta_{\mathrm{CG}}}

\graphicspath{%
    {converted_graphics/}
    {D:/himanshu/backups/backup_27Jan2007/data/paper/current_switching/lcmo18/HotElectron/TeX_PDF_Files/FigureFiles/}
}

\begin{document}
\SIunits[thickqspace]
\title{Hot electron effects and non--linear transport in hole doped manganites}
\date{\today}
\author{Himanshu Jain}
\email[E--Mail: ]{himanshu@physics.iisc.ernet.in}
\affiliation{Department of Physics, Indian Institute of Science, Bangalore 560 012, India}
\author{A. K. Raychaudhuri}
\email[E--Mail: ]{arup@bose.res.in}
\altaffiliation{On lien from: Dept of Physics, Indian Institute of Science, Bangalore 560 012, India.}
\affiliation{S. N. Bose National Centre for Basic Sciences, Salt Lake, Kolkata 700 098, India}

\begin{abstract}

We show that strong non--linear electron transport in the ferromagnetic insulating (FMI) state of manganites, responsible for phenomena such as colossal electroresistance and current induced resistance switching, can occur due to a hot electron effect. In the FMI state, which we show is an insulator with a Coulomb gap, the temperature of the electron and lattice baths can decouple at high input power levels, leading to heating of the electron bath. Parameters of the hot electron effect model were independently determined via time dependence experiments and are in good agreement with the experimental values.

\end{abstract}


\maketitle

Manganites $\mang{L}{1-x}{A}{x}$, where $\mathrm{L}$ and $\mathrm{A}$ are ions of a trivalent rare--earth and a divalent metal atom respectively, are known to exhibit a colossal magnetoresistance. Manganites also show electroresistance (ER) and related resistance state switching effects~\cite{OdagawaPRB04LiuAPL00}. There is a resurgence of interest in these materials due to the possibility of their usage in memory devices~\cite{XingAPL07} based on electric (as opposed to magnetic) control of their electrical resistance. These effects are manifestations of a strong non--linear conduction (NLC)~\cite{GuhaAPL99} in these materials, wherein their resistance is a strong function of the bias current. NLC is generally observed in manganites having hole concentration `$\mathrm{x}$' such that the ground state is insulating, e.g., in the charge and orbitally ordered insulating ground state observed at $\mathrm{x}\simeq0.5$, or in the ferromagnetic insulating (FMI) state, typically observed for $\mathrm{x}\stackrel{<}{\sim}0.3$. Though NLC phenomena for different ranges of `x' are ostensibly similar, their physical origins are likely different. In this paper, we investigate NLC in the FMI state of manganites and suggest a mechanism that may underly it. NLC in the FMI state is responsible for a ``colossal'' electroresistance (CER) and related electric current induced resistance switching (CIRS)~\cite{JainAPL06,JainPRB07}. We show that NLC and electronic properties in the FMI state (an insulator with a Coulomb--gap) are intimately related. The NLC is modeled using hot electron effects, wherein exist separate temperature scales for the electron and the phonon/lattice baths. To our knowledge, this is the first proposal that such hot electron effects occur in manganites.  

\begin{figure}[tbp] 
  \centering
  \includegraphics[bb=0 0 310 246,width=7cm,height=7cm,keepaspectratio]{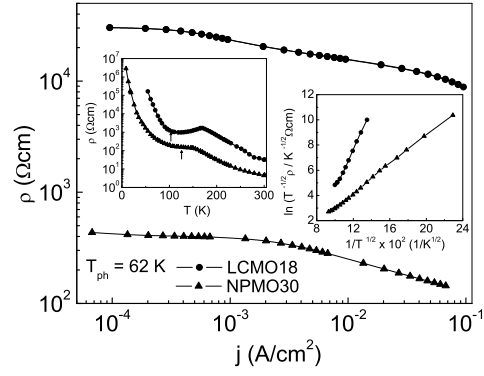}
  \caption{Representative resistivity $\rho$ vs current density $\mathrm{j}$ data at indicated fixed phonon temperature $\tph$. Left inset: Temperature $\mathrm{T}$ dependent $\rho$. The respective $\tfmi$'s are indicated by arrows. Right inset: Shklovskii Efros variable range hopping transport in the FMI state.}
  \label{fig:Fig1L18N30Rhoj62KInRhoTSEVRH}
\end{figure}

We study single crystals of two widely different manganites, $\mang{La}{0.82}{Ca}{0.18}$ (LCMO18) and $\mang{Nd}{0.7}{Pb}{0.3}$ (NPMO30). As temperature $\mathrm{T}$ is decreased, the samples undergo paramagnetic--ferromagnetic transitions at $\mathrm{T}=\tc\simeq165\usk\kelvin$ and $\simeq150\usk\kelvin$ respectively, and enter their FMI states at $\mathrm{T}=\tfmi\simeq100\usk\kelvin$ and $\simeq130\usk\kelvin$ respectively (see left inset of figure~\ref{fig:Fig1L18N30Rhoj62KInRhoTSEVRH}; the data were obtained at very low power bias). Examples of NLC data are shown in figure~\ref{fig:Fig1L18N30Rhoj62KInRhoTSEVRH}. The strong dependence of resistivity $\rho$ on current density $\mathrm{j}$ is evident. Extensive NLC data in these samples is available~\cite{JainAPL06,JainPRB07}. 

The transport in the FMI state is unique: it is activated below $\tfmi$; however, the FMI state is not a conventional band or Mott--Hubbard insulating state that have a gap in the density of states (DOS) near the Fermi level $\ef$. Low temperature heat capacity measurements of both LCMO18 and NPMO30 reveal the presence of a finite electronic contribution $\gamma\mathrm{T}$ to the heat capacity. Such a situation can arise in a Shklovskii--Efros~\cite{ShklovskiiBOOK84} insulator with a soft--gap (Coulomb--gap) $\deltacg$, and indeed, we find that $\rho$ in the FMI state has a Shklovskii Efros variable range hopping (SE--VRH) T dependence:

\begin{equation}
\label{eqn:transportmechanism}
\rho=\rho_0^\prime\exp\left(\frac{\tnot}{\mathrm{T}}\right)^{\mathrm{1/2}}
\end{equation}

where, $\rho_0^\prime=\rho_0\mathrm{T}^{1/2}$. Here, $\tnot$ represents an energy expressed in Kelvin. In right inset of figure~\ref{fig:Fig1L18N30Rhoj62KInRhoTSEVRH} we show a plot of $\mathrm{T}^{-1/2}\ln\rho$ versus $\mathrm{T}^{-1/2}$ for $\mathrm{T}\stackrel{<}{\sim}\tfmi$ to show that the $\rho$ follow equation~\ref{eqn:transportmechanism} below $\tfmi$. The experimentally estimated values of $\rho_0$, $\tnot$, and $\deltacg$ are listed in table~\ref{tab:fitparameters}. The presented values of $\deltacg$ are in agreement with direct estimation of $\deltacg$ by angle resolved photoemission spectroscopy (ARPES)~\cite{DessauS00JoyntS00}. 

\begin{figure}[tbp] 
  \centering
  \includegraphics[bb=0 0 305 313,width=7cm,height=7cm,keepaspectratio]{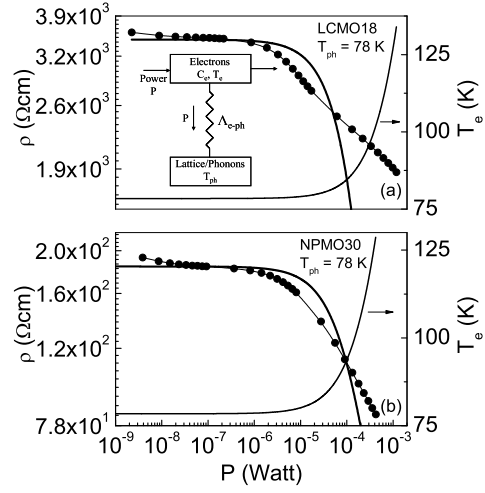}
  \caption{Experimentally observed variation of resistivity $\rho$ as a function of power $\mathrm{P}$ for (a) LCMO18, and (b) NPMO30, at indicated fixed phonon temperature $\tph$. The respective fits to the hot electron model (shown inset) are also shown. Right ordinate: calculated variation (using equation~\ref{eqn:electrontemperature}) of electron temperature $\te$ as a function of $\mathrm{P}$.}
  \label{fig:Fig2L18N30T78KRhoTevsPModel}
\end{figure}

We propose that NLC effects in manganites arise due to heating of the electrons because of the presence of a thermal conductance, that is finite, between the electron bath and the phonon bath. In this situation the electron bath temperature $\te$ will get decoupled from that of the phonon bath $\tph$ when the power $\mathrm{P}$ input is large. Direct measurements of $\tph$ have established that during measurements using higher $\mathrm{P}$, $\tph$ does not undergo any substantial rise ($\stackrel{<}{\sim}5\usk\kelvin$)~\cite{JainAPL06}, i.e., the observed NLC cannot be due to heating of the sample because of Joule dissipation. Within the present model, shown schematically in figure~\ref{fig:Fig2L18N30T78KRhoTevsPModel}, the NLC effects arise because of a heating of the electron bath. The energy exchange between the two baths is limited by an effective thermal conductance $\Lambda_\mathrm{e-ph}$, modeled empirically as $\Lambda_\mathrm{e-ph}=\Lambda_0\mathrm{T}^\alpha$. Thermal conductances generally increase as the temperature is raised, implying $\alpha\geq1$. The phonon bath, in turn, is linked  to the temperature controlled base through a finite thermal conductance. However, since it is known~\cite{JainAPL06} that $\tph$ remains close to the base temperature, we ignore the finite thermal resistance between the phonons and the base. The heat flow is thus limited by $\Lambda_\mathrm{e-ph}$. Using the relation, $\Delta\mathrm{T}\equiv(\te-\tph)=\mathrm{P}/\Lambda_\mathrm{e-ph}$, we obtain a working relation between $\tph$, $\te$, and $\mathrm{P}$:

\begin{equation}
\label{eqn:electrontemperature}
\te=\left(\tph^{\alpha+1}+\frac{\alpha+1}{\Lambda_0}\mathrm{P}\right)^{1/\alpha+1}
\end{equation}

The values of $\tnot$ and $\rhonot$ being known, at fixed $\tph$, equations~\ref{eqn:electrontemperature} and~\ref{eqn:transportmechanism} together express $\rho$ as a function of $\mathrm{P}$ in terms of only two parameters, namely $\alpha$ and $\Lambda_0$. Representative fits to this hot electron model for LCMO18 and NPMO30 at indicated $\tph$ are shown in figure~\ref{fig:Fig2L18N30T78KRhoTevsPModel}. It can be seen that the simple model based on two parameters provide good fit to the data although there are deviations in the regimes of high power. The values of $\alpha$ and $\Lambda_0$ based on the hot electron model are listed in table~\ref{tab:fitparameters}. We discuss below that these parameter values can be related to experimentally determined numbers and are completely justifiable. In figure~\ref{fig:Fig2L18N30T78KRhoTevsPModel} we also plot on the right ordinate the variation of $\te$ as a function of $\mathrm{P}$, as obtained via equation~\ref{eqn:electrontemperature}. It can be seen that at low power $\mathrm{P}<10^{-6}\usk\watt$, $\te$ is close to $\tph$ as expected, but at higher power, $\te$ increases rapidly and can be substantially greater than $\tph$ for $\mathrm{P}>10^{-3}\usk\watt$.

The model proposed here is in the line of similar effects in heavily doped semiconductors~\cite{GaleazziPRB07} with carrier concentration near the critical regime of metal--insulator transition. This regime is known to be dominated by localized electronic states and the conduction process is of the Shklovskii Efros type with a $\deltacg$ in the DOS. The key difference between the two systems is in the scale of carrier concentrations and the scale of $\deltacg$. In doped semiconductors such as Si, these effects are manifested at carrier concentrations in the range of $10^{17}$ to $10^{18}\usk\centi\power{\meter}{-3}$, and with $\deltacg\simeq1\usk\milli\electronvolt$. In the present hole doped manganites, the carrier concentrations are a few $10^{21}\usk\centi\power{\meter}{-3}$, and $\deltacg$ is about $2$ orders of magnitude larger. Due to the high value of $\deltacg$ these effects are visible in manganites at such high temperatures $\sim100\usk\kelvin$, in contrast to heavily doped semiconductors where they typically occur below $1\usk\kelvin$.

The proposed model can be validated by an independent estimation of the two parameters $\alpha$ and $\Lambda_0$ to compare with the values obtained by fitting the model to the NLC data. Reliable estimates of these parameters can be directly obtained in the following way: a relaxation time determines the energy transfer between the two baths (electron and phonon). In the temperature range we are working, the phonon bath heat capacity is much larger than that of the electronic bath. (For example, even at $50\usk\kelvin$, the electronic heat capacity $\mathrm{C_e}$ is not more than $2\%$ of the lattice heat capacity.) Thus the phonon bath acts like an infinite heat capacity bath to the electrons and in this situation, the relaxation time $\tau_\mathrm{e-ph}$ is given  by:

\begin{equation}
\label{eqn:relaxation}
\tau_\mathrm{e-ph}=\frac{\mathrm{C_e}}{\Lambda_\mathrm{e-ph}}
\end{equation}

\begin{figure}[tbp] 
  \centering
  \includegraphics[bb=0 0 374 242,width=7cm,height=7cm,keepaspectratio]{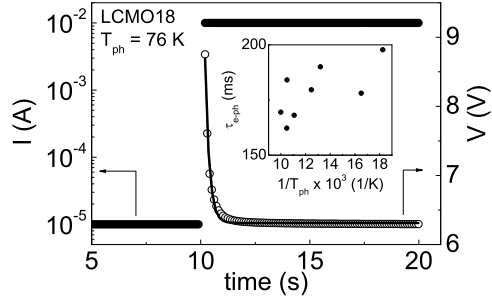}
  \caption{A typical voltage $\mathrm{V}$ relaxation profile for LCMO18 upon the application of a positive step change in current bias $\mathrm{I}$ at time $=10\usk\second$, at indicated phonon temperature $\mathrm{T_{ph}}$. The solid line is a fitted exponential decay curve used to extract the electron--phonon relaxation time constant $\tau_\mathrm{e-ph}$. Inset: $\tph$ dependence of $\tau_\mathrm{e-ph}$.}
  \label{fig:Fig3L18T76KVHiRelInTauvsTinv}
\end{figure}

Using $\mathrm{C_e}=\gamma\mathrm{T}$, one gets $\tau_\mathrm{e-ph}\approx\gamma\mathrm{T}^{1-\alpha}/\Lambda_0$. $\gamma$ is known from the heat capacity measurements. To obtain $\alpha$ and $\Lambda_0$ we need to know $\tau_\mathrm{e-ph}$, which is measured by a step change experiment as described below. At a given temperature below $\tfmi$, we increase bias current $\mathrm{I}$ by a step, causing a decoupling of $\tph$ and $\te$. Since $\te$ needs a finite time to reach equilibrium, the voltage $\mathrm{V}$ across the sample (which measures the resistance, which in turn depends on $\te$) lags behind the current step. We measure the time for $\mathrm{V}$ to relax to the new value. An example of such a voltage relaxation experiment for LCMO18 is shown in figure~\ref{fig:Fig3L18T76KVHiRelInTauvsTinv}. The voltage evolution follows a simple exponential function with a single relaxation time, which we argue is essentially the $\tau_\mathrm{e-ph}$: the fact that the $\te$ changes with a finite relaxation time when a step power is applied to the electron bath would mean that the resistance, which is function of $\te$ will evolve with time giving a finite response time for the voltage. (Note: there may be two other causes for the voltage to respond with finite time: (1) a capacitive effect, and (2) a finite thermal relaxation time for the lattice temperature to relax to the temperature controlled bath temperature, i.e., for the sample to thermalize. The first effect has been tested for and was found absent, and as for the second effect, reasonable estimates as well as measurements show that this relaxation is an order of magnitude faster than that seen in the present experiment.) Thus, the measured relaxation time is essentially $\tau_\mathrm{e-ph}$, which was found to be have a weak temperature dependence. Since $\tau_\mathrm{e-ph}\sim\mathrm{T}^{1-\alpha}$, the weak temperature dependence implies $(\alpha-1)$ to be very small, i.e., $\alpha\simeq1$. The numerical value of $\tau_\mathrm{e-ph}\simeq200\usk\milli\second$. These facts together yield an estimate for $\Lambda_0\sim2-6\times10^{-7}\usk\joule\per\kelvin\squared\second$. The values of $\alpha$ and $\Lambda_0$ determined from the fit to the data using the model (see table~\ref{tab:fitparameters}) are in good agreement with these estimates. Similar agreement was obtained for NPMO30 as well. Thus the simple model is validated by an independent experimental estimation of parameters.

We note that the electron--phonon energy relaxation which is parameterized through a single $\tau_\mathrm{e-ph}$, needs a finite $\mathrm{C_e}$, i.e., a non--zero $\gamma$. This can only happen when the insulating state is created by localization type phenomena, i.e., when the sample is not a band insulator. Presumably therefore, the type of hot electron phenomena that we are proposing would need a specific type of insulating state, and may not be applicable in other types of insulators, at least not in the present form.

The deviation  between the simple hot electron model and the experimental data at somewhat high power levels can be due to the presence of electronic inhomogeneities in the FMI state. Tacit in the simple model analysis is an assumption of uniform dissipation of power within the sample. This assumption is not strictly true for manganites, especially those having compositions close to M--I phase transition boundaries, as is the case for the compositions studied here. The inhomogeneous conductivity can lead to inhomogeneous power dissipation within the electron system, and thus to an inhomogeneous distribution of $\te$.  

To summarize, we have investigated non--linear conduction in hole doped manganites in the hole concentration region where they show a FMI state. The insulating state in the presence of ferromagnetic order in the hole doped manganites $\mang{La}{0.82}{Ca}{0.18}$ and $\mang{Nd}{0.7}{Pb}{0.3}$ is caused by electron localization due to Coulomb interaction and the charge transport is of the SE--VRH type. The value of the Coulomb--gap obtained from the data are large $\simeq10^2\usk\milli\electronvolt$. We find that a simple model of electron heating can explain the data. The increase in bias current densities, i.e., increase in input power, leads to a heating up of the electron bath which is coupled only weakly to the phonon bath via a finite thermal conductance. The parameters of the model, estimated through independent experiments, were found to be close to those obtained by fitting the model to the data.

\clearpage

\begin{acknowledgments}

HJ thanks CSIR for a fellowship and V. Ganesan of IUC Indore for making available his PPMS facility for the heat capacity measurements. AKR thanks DST for a sponsored project. HJ and AKR thank Ya. M. Mukovskii and H. L. Bhat for providing the samples.

\end{acknowledgments}

\clearpage

\begin{table}
\begin{tabular}{cccccc}\hline
Composition & $\rho_0$ & $\mathrm{T}_0$ &  $\deltacg$ & $\Lambda_0$ & $\alpha$\\
&$(\ohm\centi\meter\kelvin^{-1/2})$ & $(\kelvin)$ & $(\milli\electronvolt)$ & $(\watt\per\power{\kelvin}{1+\alpha}\meter)$ & $\one$\\\hline
LCMO18 & $5.6\times10^{-6}$ & $2.5\times10^{4}$ & $200$ & $2\times10^{-7}$ & $1$\\
NPMO30 & $1.8\times10^{-2}$ & $3.9\times10^{3}$ & $40$ & $2\times10^{-8}$ & $1.3$\\\hline
\end{tabular}
\caption{\label{tab:fitparameters}Shklovskii--Efros transport, $\rhonot$, $\tnot$, and Coulomb gap $\deltacg$, and hot electron model, $\Lambda_0$ and $\alpha$ parameters.}
\end{table}

\clearpage

\section*{List of Figure Captions}
\begin{enumerate}
\item[FIG.1:]{Representative resistivity $\rho$ vs current density $\mathrm{j}$ data at indicated fixed phonon temperature $\tph$. Left inset: Temperature $\mathrm{T}$ dependent $\rho$. The respective $\tfmi$'s are indicated by arrows. Right inset: Shklovskii Efros variable range hopping transport in the FMI state.}
\item[FIG.2:]{Experimentally observed variation of resistivity $\rho$ as a function of power $\mathrm{P}$ for (a) LCMO18, and (b) NPMO30, at indicated fixed phonon temperature $\tph$. The respective fits to the hot electron model (shown inset) are also shown. Right ordinate: calculated variation (using equation~\ref{eqn:electrontemperature}) of electron temperature $\te$ as a function of $\mathrm{P}$.}
\item[FIG.3:]{A typical voltage $\mathrm{V}$ relaxation profile for LCMO18 upon the application of a positive step change in current bias $\mathrm{I}$ at time $=10\usk\second$, at indicated phonon temperature $\mathrm{T_{ph}}$. The solid line is a fitted exponential decay curve used to extract the electron--phonon relaxation time constant $\tau_\mathrm{e-ph}$. Inset: $\tph$ dependence of $\tau_\mathrm{e-ph}$.}
\end{enumerate}

\clearpage

\clearpage

\clearpage

\end{document}